\documentclass[11pt]{article}
\usepackage{latexsym}
\usepackage[dvips]{graphics,color}
\usepackage{graphpap}
\usepackage{alltt}
\usepackage{graphicx}
\usepackage{amsfonts}
\usepackage{amsmath}
\usepackage{psfrag}
\newtheorem{theorem}{Theorem}[section]

\newtheorem{proposition}[theorem]{Proposition}

\newtheorem{definition}[theorem]{Definition}

\newcommand{\beq}{\begin{equation}}
\newcommand{\feq}[1]{\label{#1} \end{equation}}
\newcommand{\beqr}{\begin{eqnarray}}
\newcommand{\feqr}{\end{eqnarray}}
\def\non{\nonumber}

\newcommand{\rf}[1]{(\ref{#1})}
\def\pr{^{\prime}}
\DeclareFontFamily{U}{eufm}{}
\DeclareFontShape{U}{eufm}{m}{n}{<->eufm10}{}
\DeclareSymbolFont{mcy}{U}{eufm}{m}{n}
\DeclareMathSymbol{\Hr}{\mathord}{mcy}{"58}

\def\np#1#2#3{Nucl. Phys. {\bf{B#1}} (#2) #3}
\def\cmp#1#2#3{Comm. Math. Phys. {\bf{#1}} (#2) #3}
\def\bul#1#2#3{Bull. London Math. Soc. {\bf{#1}} (#2) #3}
\def\plms#1#2#3{Proc. London Math. Soc. {\bf{#1}} (#2) #3}
\def\pr#1#2#3{Phys. Rep. {\bf{#1}} (#2) #3}

\def\am#1#2#3{Ann. of Math. {\bf{#1}} (#2) #3}

\def\jfa#1#2#3{J. Funct. Anal. {\bf{#1}} (#2) #3}

\def\ase#1#2#3{Ann. Scient. \'{E}c. Norm. Sup. {\bf{#1}} (#2) #3}

\def\jhp#1#2#3{JHEP {\bf{#1}} (#2) #3}

\usepackage[active]{srcltx}
\begin{document}


\begin{center}


{\Large \bf One-loop Partition Functions in Hyperbolic Space $\mathbb{H}^n(\mathbb{R})$}\\
[4mm]

\large{Agapitos N. Hatzinikitas} \\ [5mm]

{\small Department of Mathematics, \\ 
University of Aegean, \\
School of Sciences, \\
Karlovasi, 83200\\
Samos Greece \\
E-mail: ahatz@aegean.gr}\\ [5mm]

\end{center}

\begin{abstract}
We first study the problem of the one-loop partition function for a free massive quantum field theory living on a fixed background hyperbolic space on the field of real numbers, $\mathbb{H}^n(\mathbb{R}), \,\, n\geq 2$. Earlier attempts were limited to $n=3$ dimensions due to the computational complexity. We have developed a new method to determine the fundamental solution of the heat equation and techniques to specify its asymptotics in the small time limit. These enable us to determine the regular part and the ultra violet divergences of the one-loop effective action in the scalar case. The contribution of the Abelian gauge excitations to the one-loop partition function were treated separately using Fourier analysis and bi-tensor techniques on $\mathbb{H}^n(\mathbb{R})$. Finally, by employing the DeWitt's method, we confirmed the correctness and extended our results to any dimension regarding the first three heat kernel coefficients.
\end{abstract}

\noindent\textit{PACS 2010:} 04.62.+v, 11.10.-z, 11.15.-q, 02.30.-f, 02.40.Ky  
\section{Introduction}
\label{sec0}
In this paper our goal is to determine explicitly  the Euclidean partition function 
\beqr
Z=\int_{\mathbb{H}^n} \mathcal{D}\phi \, e^{-g^{-2}S_E(\phi)}
\label{sec0 : eq1}
\feqr
of a free quantum scalar field $\phi$ and of a $U(1)$ gauge quantum field. The factor $g^{-2}$ included in front of the Euclidean action is proportional to $1/\hbar$. This problem along with the graviton excitations were solved in three dimensions by \cite{GM}. We pursuit the more general $n$-dimensional case for mathematical completeness.     
\par If we partial integrate the corresponding actions of the fields, by imposing suitable boundary conditions, we arrive at 
\beqr
S_E(\phi)&=&\frac{1}{2}\left(\langle d\phi,d\phi\rangle+m^2\langle \phi,\phi\rangle\right)=\frac{1}{2}\int_{\mathbb{H}^n}\phi \hat{D}_{\phi}\phi \, \sqrt{g}d^nx, \quad \hat{D}_{\phi}=-\nabla^{\mu}\partial_{\mu}+m^2
\label{sec0 : eq2} \\
S_E(A,c,b)&=&\frac{1}{2}\left(\langle F,F\rangle +\langle \star d \star A,\star d\star A\rangle+ m^2\langle A,A\rangle\right)+\langle dc,db\rangle \nonumber \\
&=& \frac{1}{2}\int_{\mathbb{H}^n}A^{\mu} \hat{D}_{F,\mu \nu}A^{\nu} \, \sqrt{g}d^nx+\int_{\mathbb{H}^n}c\hat{D}_{gh} b \, \sqrt{g} d^n x, \nonumber \\
\hat{D}_{F,\mu\nu}&=& g_{\mu \nu}(-\Delta+m^2) - R_{\mu \nu},\quad \hat{D}_{gh}=-\nabla^{\mu}\partial_{\mu}
\label{sec0 : eq3}
\feqr  
where $\langle \cdot, \cdot \rangle$ is the Hodge inner product (see \rf{apA :  eq1a} for the definition), $\star$ is the Hodge star operation and $d$ the exterior derivative \cite{F}, \cite{CH}. In \rf{sec0 : eq3} we adopted the Feynman gauge, to simplify the form of the operator $\hat{D}_{F,\mu\nu}$, and the Fadeev-Popov ghosts $c, b$ have been introduced by the gauged fixed Yang-Mills action.  
\par Since we are dealing with determinants of operators, it is instructive to provide some information about their structure. The operator $\hat{D}_{\phi}$ is semi-bounded from below by $\lambda_0=(n-1)^2/4+m^2$ and symmetric on its domain $dom(C_0^{\infty}(\mathbb{H}^n))$. If $\hat{D}_{\phi}$ has a self-adjoint closure, denoted by $\overline{\hat{D}}_{\phi}$, then it is said to be an essentially self-adjoint operator \cite{RS}. Since the hyperbolic manifold is geodesically complete, such self-adjoint extension is uniquely determined \cite{GA}, \cite{R}, \cite{S} and the problem of defining the Feynman propagator (i.e., the Green's operator $\hat{D}_{\phi}^{-1}$) can naturally be solved. In the remainder of the paper we will consider $\overline{\hat{D}}_{\phi}$ and from now on the bar will be omitted from the extension of $\hat{D}_{\phi}$. Negative powers of $\hat{D}_{\phi}$ can be expressed in terms of the heat operator $e^{-t \hat{D}_{\phi}}, \, t>0$ by the following integral
\beqr
\hat{D}_{\phi}^{-l}= \frac{g(l)}{\Gamma(l)}, \quad \textrm{where} \quad 
g(l)=\int_0^{\infty}e^{-t \hat{D}_{\phi}}t^{l-1} dt
\label{sec0 eq31}
\feqr
and $\ln(\hat{D}_{\phi})$ by
\beqr
\ln(\hat{D}_{\phi})=-\lim_{\epsilon\rightarrow 0^+}\left(\int_{\epsilon}^{\infty}\frac{e^{-t\hat{D}_{\phi}}}{t} dt+(\gamma+\ln \epsilon)\hat{I}\right)
\label{sec0 eq32}
\feqr
where $\hat{I}$ denotes the identity operator and $\gamma$ is Euler's or Mascheroni's constant 
\beqr
\gamma=-\int_0^{\infty}e^{-s}\ln s\,  ds.
\label{sec0 eq33} 
\feqr
The multiple of the identity operator in \rf{sec0 eq32} will be discarded from now on since we are considering renormalized effective actions. 
\par Consider the Hilbert space $\mathcal{H}=\mathcal{L}^2(\mathbb{H}^n,<\cdot,\cdot>)$ with inner product defined by
\beqr
<u,v>=\int_{\mathbb{H}^n}u(x)v(x)\, \sqrt{g}d^nx
\label{sec0 eq34}
\feqr
for real one-component scalar fields.
Adopting Dirac's notation, consider the abstract basis $|x'>\in\mathcal{H}$ and the dual basis $<x|\in\mathcal{H}^*$, then the orthonormal condition reads $<x|x'>=\delta^{(n)}(x,x')$. The heat kernel $e^{-\tau\hat{D}_{\phi}}$ has elements
\beqr
K_n(x,x';\tau)=<x|e^{-t\hat{D}_{\phi}}|x'>.
\label{sec0 eq35}
\feqr 
\par The contribution of quantum effects to the one-loop partition function in the scalar field case is given by, \cite{VA} (and references therein), the functional
\beqr
W^{(1-loop)}=-\frac{1}{2}\ln\textrm{det}(\hat{D}_{\phi})=-\frac{1}{2}\textrm{Tr}\ln(\hat{D}_{\phi}).
\label{sec0 eq4}
\feqr  
If the manifold $\mathcal{M}$ were compact then $\hat{D}_{\phi}$ would have a discrete spectrum and \rf{sec0 eq4} with the help of \rf{sec0 eq32} would become
\beqr
W^{(1-loop)}= \frac{1}{2} \lim_{\epsilon\rightarrow 0^+}\int_{\epsilon}^{\infty} \frac{dt}{t} \textrm{Tr}\left(e^{-t \hat{D}_{\phi}} \right)
=\frac{1}{2} \lim_{\epsilon\rightarrow 0^+}\int_{\epsilon}^{\infty} \frac{dt}{t}\int_{\mathcal{M}}K_n(x,x;t) \sqrt{g} d^n x.
\label{sec0 eq6}
\feqr
The hyperbolic space is not compact and $\hat{D}_{\phi}$ would also have a continuous spectrum giving a divergent contribution to \rf{sec0 eq6} proportional to the volume of $\mathcal{M}$. 
\par Relation \rf{sec0 eq6} can be calculated by evaluating the heat kernel at the coincident limit $x'\rightarrow x$. Also, due to the presence of the mass term, one is not  confronted with infra-red (IR) divergences caused by zero or negative eigenvalues of $\hat{D}_{\phi}$. Only ultra-violet (UV) divergences are present. These can be found by introducing a lower cut off $\Lambda^{-2}$ and performing the $\Lambda\rightarrow \infty$ limit. For $\mathbb{H}^{2k+1}$ the fundamental solution of the heat equation is expressed in terms of elementary functions and this allows us to obtain an explicit formula for the one-loop effective action. In even dimensions the analysis is more involved but the recurrence relation \rf{sec51 : eq7b} enables us to solve this problem.      
\par Regarding the Abelian gauge excitations, the operator $\hat{D}_{F,\mu\nu}$ is semi-bounded from below by $\lambda_0=(n-3)^2/4+m^2-(n-1)$ and symmetric on its domain $\textrm{dom}(C_0^{\infty}(\mathbb{H}^n, \mathfrak{u}(1)))$ \footnote{$\mathfrak{u}(1)$ is the unitary Lie Algebra.}. One can repeat operator theory arguments, as the ones developed in the scalar case, to establish  essential self-adjointness of $\hat{D}_{F,\mu\nu}$. The one-loop partition function now becomes
\beqr
W^{(1-loop)}_{U(1)}&=&-\frac{1}{2}\left(\ln\textrm{det}(\hat{D}_{F,\mu\nu})-2\ln\textrm{det}(\hat{D}_{gh})\right) \nonumber \\
&=&\frac{Vol(\mathbb{H}^{n})}{2}\lim_{\epsilon\rightarrow 0^+}\int_{\epsilon}^{\infty}\frac{dt}{t}\left(\textrm{Tr}(K_{U(1)}^{\mathbb{H}^{n}})(t)-K^{\mathbb{H}^{n}}_{gh}(t)\right).
\label{sec0 eq7}
\feqr
where the factor $-2$ is due to the presence of $b, c$ fields. The vector nature of the $U(1)$ Abelian gauge field implies a delicate treatment for the one-loop effective action. Adopting a special ansatz for the heat kernel, first proposed by \cite{HF}, and using Fourier analysis as well as bi-tensor identities on $\mathbb{H}^n$, we reach our goal. Motivated by the DeWitt's powerful method, after some modifications, we performed an independent calculation of the heat kernel coefficients. This method produced identical results and extended the first ones.         
\par The paper is organized as follows:
\par In \textit{Section 2} we briefly review the definition of the real $n-$dimensional hyperbolic space $\mathbb{H}^n$ and list some fundamental properties.
\par In \textit{Section 3}, depending on the dimension of the manifold $\mathbb{H}^n$ (even or odd), we derive a recurrent relation for the fundamental solution of the heat equation in terms of the probability density functions \rf{sec3 : eq1ab} and \rf{sec3 : eq1b} respectively. The method is new and elegant and deserves a special mention. In sections five and six these recurrent relations will be used repeatedly.  
\par In \textit{Section 4} we present the Fourier transform on $\mathbb{H}^n$ dictated by a version of Helgason's treatment who involves group theoretic machinery. We write the unique $SO(n)$ invariant spherical eigenfunction $\Phi_{\lambda}(x)$ of the Laplace-Beltrami operator subjected to the condition $\Phi_{\lambda}(0)=1$. This help us to identify the Harish-Chandra $c-$function (Proposition 4.2) which plays a key role in harmonic analysis on $\mathbb{H}^n$. We state the basic theorem without proof for the Fourier transformation on symmetric spaces of non-compact type and concentrate on rotationally symmetric functions.
\par In \textit{Section 5} we determine the one-loop partition function of a massive and real scalar quantum field theory. The regular and divergent parts are calculated in any odd, two and four dimensions respectively. A different approach based on the W.K.B. approximation of the heat kernel is also performed to secure our results. The latter provides results independent of the dimension of the hyperbolic space. We also derive a weighted Poincar\'{e} inequality for radial functions.   
\par In \textit{Section 6} we solve the same problem as previously but in the presence of a massive Abelian $U(1)$ gauge field. Adopting a convenient ansatz for the $U(1)$ heat kernel and using Fourier analysis we specify the trace of the corresponding kernel. Again an independent W.K.B. heat kernel approximation is performed to check the agreement of the first three heat kernel coefficients with those predicted by \rf{sec0 eq7}. 
\par In \textit{Section 7} we provide additional mathematical background to clarify technical issues.   
\section{Background}
\label{sec1}
\subsection{Definition of $\mathbb{H}^n(\mathbb{R})$ and properties}
\label{sec1a}
\begin{definition}
The hyperbolic space $\mathbb{H}^n, \, n\geq 2$ geometrically can be realised as the upper sheet of the hyperboloid embedded into the Minkowski space, $\mathcal{M}^{(1,n)}$, namely
\beqr
\mathbb{H}^n=\{X\in \mathcal{M}^{(1,n)} :\,\, [X,X]=1, \, X_{0}>0\}
\label{sec1 : eq1}
\feqr
where $[\cdot,\cdot]$ is the bilinear form defined by
\begin{displaymath}
[X,Y]=X_0Y_0-\sum_{i=1}^{n}X_iY_i=X^tJ_{1,n}Y
\end{displaymath}
with $J_{1,n}$ be the $(n+1)\times(n+1)$ diagonal matrix with signature $(+1,-1,\cdots,-1)$.
\end{definition}
The hyperbolic space $\mathbb{H}^n$ is the maximally symmetric, simply connected, $n$-dimensional Riemannian manifold with constant negative sectional curvature $K=-1$. The group of all orientation preserving isometries, $\textrm{Iso}^+ (\mathbb{H}^n)$, consists of the proper, orthochronous Lorentz group 
\beqr
G&=&\textrm{Iso}^+ (\mathbb{H}^n)=SO^+(1,n)=SO(1,n)\cap O^+(1,n) \nonumber \\
&=&\{GL(n+1,\mathbb{R}): \,\, h^tJ_{1,n}h=J_{1,n}, \, \textrm{det}h=1, \, h_{00}>0\}
\label{sec1 : eq2}
\feqr
where $GL(n+1,\mathbb{R})$ is the group of all nonsingular real $(n+1)\times (n+1)$ matrices $h$. This group is isomorphic to the M\"{o}bius group of $\overline{\mathbb{R}}^{n-1}=\mathbb{R}^{n-1}\cup \{\infty\}$. An element of $h\in G$ acts on $x\in \mathbb{H}^n$ as follows
\beqr
x_i'&=&\frac{(h_{i0}+h_{in})|x|^2 +2\sum_{k=1}^{n-1}h_{ik}x_k+h_{i0}-h_{in}}{A_h|x|^2+2\sum_{k=1}^{n-1}(h_{0k}-h_{nk})x_k+B_h} \quad i=1,\cdots,n-1
\label{sec1 : eq3a} \\
x_n'&=&\frac{2x_n}{A_h|x|^2+2\sum_{k=1}^{n-1}(h_{0k}-h_{nk})x_k+B_h}
\label{sec1 : eq3b}
\feqr
where $|x|^2=\sum_{k=1}^{n}x_k^2$, $A_h=h_{00}+h_{0n}-h_{n0}-h_{nn}$ and $B_h=h_{00}+h_{nn}-h_{n0}-h_{0n}$.
A metric on $\mathbb{H}^n$ can be constructed by  
\begin{displaymath}
g_{\mathcal{M}^{(1,n)}}=\sum_{i=1}^{n}\left(dX_i\right)^2-\left(dX_{0}\right)^2.
\end{displaymath}

\subsection{Heat kernel recurrence relations between different dimensions}
\label{sec3}  
\begin{definition}
A fundamental solution $p_n$ for the heat operator $\partial/\partial t - \Delta_{\mathbb{H}^n,x}$ is a function 
\begin{displaymath}
p_n: \,\,\mathbb{H}^n\times \mathbb{H}^n \times (0,\infty)\rightarrow \mathbb{R}
\end{displaymath} 
with the following properties
\begin{itemize}
\item[$i)$] $p_n\in \mathcal{C}(\mathbb{H}^n\times \mathbb{H}^n \times (0,\infty))$, $\mathcal{C}^2$ in the first variable and $\mathcal{C}^1$ in the second variable,
\item[$ii)$] $\left(\partial_t-\Delta_{\mathbb{H}^n,x}\right) p_n(x,y;t)=0, \, \, \forall t>0$,
\item[$iii)$] $\lim_{t\rightarrow 0^+}p_n(\cdot,y;t)=\delta_y(\cdot), \,\, \forall y\in\mathbb{H}^n $
where $\delta_y$ is the Dirac distribution centered at $y$ and the limit is considered in the distributional sense,
\beqr
\lim_{t\rightarrow 0^+}\int_{\mathbb{H}^n}p_n(x,y;t)f(y)d^ny=f(x), \,\, \forall f\in\mathcal{C}^{\infty}_0(\mathbb{H}^n), \, \forall y\in \mathbb{H}^n
\label{sec3 : eq1a}
\feqr
where $\mathcal{C}^{\infty}_0(\mathbb{H}^n)$ denotes the set of $\mathcal{C}^{\infty}-$ smooth functions with compact support and $d^ny=\sqrt{g}\, dy^1\wedge \cdots \wedge dy^n$ is the volume form.
\end{itemize}
\end{definition}
\begin{theorem}
Let
\beqr
p_1(r,t)&=& \frac{1}{\sqrt{4\pi t}}e^{-\frac{r^2}{4t}} \label{sec3 : eq1ab} \\
p_2(r,t)&=&\frac{\sqrt{2}}{(4\pi t)^{\frac{3}{2}}}e^{-\frac{t}{4}}\int_{r}^{\infty}\frac{se^{-\frac{s^2}{4t}}}{\sqrt{\cosh s -\cosh r}} ds
\label{sec3 : eq1b}
\feqr
be the fundamental solutions for the heat operator on the real line and the hyperbolic space $\mathbb{H}^2$ respectively. Then the following recurrence relations hold 
\beqr
p_{2k+1}(r,t)&=& \frac{(-1)^k}{(2\pi)^k}e^{-k^2 t}\left(\frac{1}{\sinh r}\partial_r\right)^k p_1(r,t)
\label{sec3 : eq1c} \\
p_{2(k+1)}(r,t)&=& \frac{(-1)^k}{(2\pi)^k} e^{-k(k+1) t}\left(\frac{1}{\sinh r}\partial_r\right)^k p_2(r,t)
\label{sec3 : eq1d}
\feqr 
depending on whether $\textrm{dim} \mathbb{H}^n=n$ is an odd or even natural number.
\end{theorem}
The formulas \rf{sec3 : eq1c} and \rf{sec3 : eq1d} are not new \cite{GRI}, \cite{DAV} but the proposed method is inspired by a different viewpoint.\\
\textbf{Proof}\\
The Laplacian commutes with the action of the isometry group $G=SO^+(1,n)$ on $\mathbb{H}^n$ space and as a result $p_n$ is a point-pair invariant; that is, it satisfies 
\beqr
p_n(x,y;t)=p_n(hx,hy;t), \quad \forall x,y\in \mathbb{H}^n, \, \forall h\in G, \, t>0.
\label{sec3 : eq1}
\feqr
Therefore 
\beqr
p_n(x,y;t)=p_n(d_{\mathbb{H}^n}(x,y);t)=p_n(r,t)
\label{sec3 : eq1ba}
\feqr
where $r(x,y)=d_{\mathbb{H}^n}(x,y)$ is the geodesic distance between the points $x,y \, \in \mathbb{H}^n$.
A straightforward calculation shows that (see Appendix \rf{apA : eq4d} for the generalization)
\beqr
\Delta_x d_{\mathbb{H}^n}(x,y)=(n-1)\coth (d_{\mathbb{H}^n}(x,y))
\label{sec3 : eq1ca}
\feqr
and the diffusion equation turns out to be
\beqr
\partial_t p_n(r,t)=\partial_r^2 p_n(r,t)+(n-1)\coth r \partial_r p_n(r,t).
\label{sec3 : eq1da}
\feqr
We want to establish a recurrence relation between $p_m$ and $p_n, \, m>n$ in different dimensions of hyperbolic spaces. Suppose there exists a differential operator $\hat{B}(r,t)$ such that
\beqr
p_m(r,t)=\hat{B}(r,t)p_n(r,t)=f(r,t)\partial_r p_n(r,t).
\label{sec3 : eq2}
\feqr
Then, by demanding the heat operator to commute with the $\hat{B}$ operator,  
\beqr
\left[\partial/\partial t-\Delta_{\mathbb{H}^m},\hat{B}(r,t)\right]=0
\label{sec3 : eq2a}
\feqr
we obtain the following system of partial differential equations 
\beqr
\partial_t f-\partial_r^2 f-(m-1)\coth r \,\partial_r f -\frac{(n-1)}{\sinh^2 r}f &=&0 
\label{sec3 : eq3a} \\
2\partial_r f+(m-n)\coth r f &=&0. 
\label{sec3 : eq3b}
\feqr
Eliminating the partial derivatives with respect to $r$, from equation \rf{sec3 : eq3a} using \rf{sec3 : eq3b}, we end up with a constraint  which relates the dimensions of the hyperbolic spaces and a first order, w.r.t. time $t$, partial differential equation 
\beqr
\textrm{dim} \mathbb{H}^{m}=\textrm{dim} \mathbb{H}^{n}+2
\label{sec3 : eq4} \\
\partial_t f +n f=0.
\label{sec3 : eq4a}
\feqr 
The general solution of \rf{sec3 : eq4a} is $f(r,t)=exp(-nt)g(r)$. Substituting this back into \rf{sec3 : eq3b} we determine $g(r)$. The operator $\hat{B}$ is found to be given, up to a constant, by 
\beqr
\hat{B}(r,t)\propto e^{-nt}\frac{1}{\sinh r} \frac{\partial}{\partial r}.
\label{sec3 : eq5}
\feqr
Applying successively $\hat{B}$ we recover \rf{sec3 : eq1c} and \rf{sec3 : eq1d} with the help of the finite sum formulas $\sum_{l=0}^{k-1}(2l+1)=k^2$ and $\sum_{l=0}^{k-1}2(l+1)=k(k+1)$. 

\subsection{Fourier transform on $\mathbb{H}^n(\mathbb{R})$}
\label{sec4}

Consider the boundary cone of $\mathbb{H}^n$
\beqr
\mathcal{A}=\{X\in \mathbb{R}^n : [X,X]=0, \, X_{n+1}>0\}
\label{sec4 : eq1}
\feqr
and the mapping 
\beqr
g : \mathbb{S}^{n-1}\rightarrow \mathcal{B} 
\label{sec4 : eq2}
\feqr
with $g(\omega)=(1,\omega)$ and
\beqr
\mathcal{B}=\{X\in \mathcal{A} : X_{n+1}=1\}.
\label{sec4 : eq3}
\feqr
\begin{proposition}
The functions 
\beqr
h_{\lambda,\omega}(x)= [x,g(\omega)]^{i\lambda-\rho}, \quad \rho=\frac{n-1}{2} 
\label{sec4 : eq4}
\feqr
are eigenfunctions of the Laplace-Beltrami operator $\Delta_{\mathbb{H}^n}$ with eigenvalues $-(\lambda^2+\rho^2)$.
\end{proposition}
\textbf{Proof}\\
Using the spherical polar representation, the bilinear form is written as
\beqr
[x,g(\omega)]^{s}&=&[(\cosh r,\phi),(1,\omega)]^{s}, \,\, \phi, \omega \in \mathbb{S}^{n-1} \nonumber \\
&=&(\cosh r-\sinh r \langle\phi,\omega\rangle)^{s}\nonumber \\
&=&(\cosh r-\sinh r \cos \theta)^{s}.
\label{sec4 : eq5}
\feqr
The Laplace-Beltrami operator when expressed in the geodesic polar representation can be applied on $h$ to give
\beqr
\Delta_r h(r,\theta)=s(s-1+n)h-\frac{1}{\sinh^2 r}\Delta_{\mathbb{S}_{n-1}}h.
\label{sec4 : eq6}
\feqr
Setting $s=i\lambda -\rho$, in \rf{sec4 : eq6}, we recover the desired result. Note that for $s=(n-1)^2/4$ the $\mathcal{L}^2$ spectrum of $\hat{D}_{\phi}$ is the whole of the interval
\beqr
\textrm{Spec}(-\hat{D}_{\phi})=\bigl[ \frac{(n-1)^2}{4},\infty \bigr).
\label{sec4 : eq6a}
\feqr 
For an alternative approach the reader may refer to \cite{GR1}.
\par The spherical function \cite{OB}
\beqr
\Phi_{\lambda}(x)&=& \frac{1}{\omega_{n-1}}\int_{\mathbb{S}^{n-1}}[x,g(\omega)]^{i\lambda-\rho}d\omega, \,\, \omega_{n-1}=|\mathbb{S}^{n-1}|=\frac{2\pi^{\frac{n}{2}}}{\Gamma(\frac{n}{2})} \nonumber \\
&=& \frac{1}{\sqrt{\pi}} \frac{\Gamma\left( \rho+\frac{1}{2}\right)}{\Gamma(\rho)} \int_0^{\pi}(\cosh r-\cos \theta \sinh r)^{i\lambda -\rho}(\sin \theta)^{2\rho -1}d\theta, \quad \rho\geq 1 
\label{sec4 : eq7}
\feqr
is the unique $SO(n)$ invariant eigenfunction of $\Delta_{\mathbb{H}^n}$ satisfying the condition $\Phi_{\lambda}(0)=1$. It is an even function w.r.t. $\lambda$ and $r$. For calculational purposes relation \rf{apB : eq4c} is more convenient. Motivated by \cite{HEL} we are led to the following proposition.
\begin{proposition}
Let $Re(i\lambda)>0$ then 
\beqr
\lim_{r\to \infty}e^{-(i\lambda-\rho)r}\Phi_{\lambda}(r)=c(\lambda), \quad \textit{where} \quad c(\lambda)=\frac{2^{2\rho-1}\Gamma(\rho+\frac{1}{2})\Gamma(i\lambda)}{\sqrt{\pi}\Gamma(\rho+i\lambda)}
\label{sec4 : eq8}
\feqr
is the Harish-Chandra $c$-function.
\end{proposition} 
\textbf{Proof}\\
Using the transformation $u=\tan(\theta/2)$ the integral \rf{sec4 : eq7} becomes
\beqr
\Phi_{\lambda}(r)=e^{(i\lambda-\rho)r}\frac{2^{2\rho}}{\sqrt{\pi}} \frac{\Gamma\left( \rho+\frac{1}{2}\right)}{\Gamma(\rho)} \int_0^{\infty}(1+e^{-2r}u^2)^{i\lambda -\rho}\frac{u^{2\rho-1}}{(1+u^2)^{i\lambda+\rho}}du.
\label{sec4 : eq9a}
\feqr
Assuming $\lambda=a+ib$ and $Re(i\lambda)=-b>0$ the integrand is absolutely integrable since for $\rho\leq |b|$ we have
\beqr
|f(r,u)|=\left|(1+e^{-2r}u^2)^{ia}(1+u^2)^{ia}\right|(1+e^{-2r}u^2)^{|b|-\rho}(1+u^2)^{-|b|-\rho}u^{2\rho-1}\leq (1+u^2)^{-2\rho}u^{2\rho-1}.
\label{sec4 : eq9b}
\feqr
Note that by definition $\rho\geq 1/2$. If $\rho > |b|$ then 
\beqr
|f(r,u)|<(1+u^2)^{-|b|-\rho}u^{2\rho-1}
\label{sec4 : eq9c}
\feqr
which is again integrable. Therefore by applying the dominated convergence theorem and changing once more variable, by setting $t=(1+u^2)^{-1}$, we arrive at the following pointwise estimate
\beqr
\lim_{r\to \infty}e^{-(i\lambda-\rho)r}\Phi_{\lambda}(r)&=&\frac{2^{2\rho-1}}{\sqrt{\pi}}\frac{\Gamma\left( \rho+\frac{1}{2}\right)}{\Gamma(\rho)}\int_0^1 (1-t)^{\rho-1}t^{i\lambda-1}dt \nonumber \\
&=&c(\lambda).
\label{sec4 : eq9d}
\feqr
In \rf{sec4 : eq9d} we used the definition of the Beta function $B(\rho,i\lambda)=\Gamma(\rho)\Gamma(i\lambda)/\Gamma(\rho+i\lambda)$.  
\par Let now the Fourier transform of $f\in C^{\infty}_{0}(\mathbb{H}^n)$ be defined by
\beqr
\tilde{f}(\lambda,\omega)=\int_{\mathbb{H}^n}f(x)h_{\lambda,\omega}(x)dx.
\label{sec4 : eq9}
\feqr
On symmetric spaces of non-compact type, Helgason \cite{HEL} has given the following Theorem of Fourier transform.
\begin{theorem}
Let $f\in C^{\infty}_{0}(\mathbb{H}^n)$. Then
\begin{enumerate}
\item[($i$)]  The inverse Fourier transform is given by
\beqr
f(x)=\frac{2^{2\rho}}{2\pi \omega_{n-1}}\int_{\mathbb{R}^+}\left(P_{\lambda}f\right)(x)\frac{d\lambda}{|c(\lambda)|^2}\quad \textit{with}\quad \left(P_{\lambda}f\right)(x)=\int_{\mathbb{S}^{n-1}}\tilde{f}(\lambda,\omega)\bar{h}_{\lambda,\omega}(x)d\omega.
\label{sec4 : eq10}
\feqr
\item[($ii$)] Plancherel
\beqr
\parallel f\parallel^2_2=\frac{2^{2\rho}}{2\pi \omega_{n-1}}\int_{\mathbb{R}^+}\left|\frac{\left(P_{\lambda}f\right)(x)}{c(\lambda)} \right|^2 d\lambda
\label{sec4 : eq11}
\feqr
and the Fourier transform extends as an isometry of $\mathcal{L}^2(\mathbb{H}^n)$ functions onto 
\begin{displaymath}
\mathcal{L}^2(\mathbb{R}^+ \times \mathbb{S}^{n-1},\frac{2^{\rho}}{\sqrt{2\pi \omega_{n-1}}}|c(\lambda)|^{-2}d\omega d\lambda).
\end{displaymath}
\end{enumerate}
\end{theorem}
\par For rotationally symmetric functions the Fourier transform extends as an isometry of radial $\mathcal{L}^2(\mathbb{H}^n)$ functions onto $\mathcal{L}^2(\mathbb{R}^+,\frac{2^{2\rho}}{2\pi \omega_{n-1}}|c(\lambda)|^{-2}d\lambda)$. Note that the Harish-Chandra measure $\frac{2^{2\rho}}{2\pi \omega_{n-1}|c(\lambda)|^2}$ corresponds to the density of zero angular momentum radial functions \cite{B}. Relation \rf{sec4 : eq9} becomes 
\beqr
\tilde{f}(\lambda)=\omega_{n-1}\int_{0}^{\infty}f(r)\Phi_{\lambda}(r)(\sinh r)^{2\rho} dr
\label{sec4 : eq12}
\feqr
and the inverse transform is then given by
\beqr
f(r)=\frac{2^{2\rho}}{2\pi \omega_{n-1}}\int_{\mathbb{R}^+}\tilde{f}(\lambda)\Phi_{\lambda}(r)\frac{d\lambda}{|c(\lambda)|^2}.
\label{sec4 : eq13}
\feqr

\section{One-loop determinants in $\mathbb{H}^n(\mathbb{R})$}
\label{sec5}
\subsection{Scalar fields}
\label{sec51}

We consider a real scalar field $\phi$ of mass $m$ on $\mathbb{H}^n$. Its action is described by \rf{sec0 : eq2} where Green's formula has been applied on the condition that either $\textrm{supp$\phi$}$ or $\textrm{supp$\nabla \phi$}$ is compact. The scalar heat kernel on $\mathbb{H}^n$ has been determined by Theorem 3.2, apart from a multiplicative constant $exp(-m^2 t)$ due now to the presence of the mass term.
\par The quantum effects generated by the background field $\phi$ in the one-loop approximation of quantum field theory are given by the functional \rf{sec0 eq6} and taking into account that $K_n(x,x;t)=p_n(0,t)$ it can be rewritten as 
\beqr
W^{(1-loop)}= \frac{1}{2}Vol(\mathbb{H}^n)\lim_{\epsilon\rightarrow 0^+}\int_{\epsilon}^{\infty}\frac{dt}{t}K_n(0,t)
\label{sec51 : eq2b}
\feqr
where the volume of $\mathbb{H}^n$ is set to be equal to the volume of the open ball $B(o,r)$\footnote{Actually, in order to remove this volume divergence one may replace $\hat{D}_{\phi}$ in \rf{sec0 eq4} by $\hat{D}_{\phi}/\hat{D}_{0,\phi}$ where $\hat{D}_{0,\phi}\equiv -\partial_{\mu}^2+m^2$ is a reference operator describing the propagation of a massive particle in Euclidean space \cite{VA}. Alternatively, the regularisation of this divergence can be performed by introducing, under the trace in \rf{sec0 eq4}, a smooth smearing function $f$ defined on $\mathbb{H}^n$.}
\beqr
Vol(B)(r)=|\mathbb{S}^{n-1}|\int_{0}^{r}(\sinh w)^{n-1} dw.
\label{sec51 : eq3}
\feqr 
The asymptotic behaviour of the heat kernel for $n=2k+1$, in the $x'\rightarrow x$ coincident limit, or equivalently, using the spherical polar representation, in the $r\equiv d_{\mathbb{H}^n}(x,x')\rightarrow 0$ limit, can be written as the finite $t-$series 
\beqr
K_{2k+1}(0,t)=\frac{1}{(4\pi t)^{k+\frac{1}{2}}}e^{-(k^2+m^2)t}\sum_{l=0}^{k-1}t^l a_{k,l}.
\label{sec51 : eq4}
\feqr
The truncation of the series at the finite positive integer $k-1$ is due to the fact that the coefficients $a_{k,l}$ vanish for $l\geq k$. This can be proved by taking the $r\rightarrow 0^+$ limit of the multiple derivative \rf{apB : eq4b0} (see Appendix B for some analytic expressions of $p_{2k+1}$). Substituting \rf{sec51 : eq4} into \rf{sec51 : eq2b} we obtain
\beqr
W^{(1-loop)}_{2k+1, reg.}=\frac{(-1)^{k+1}}{2^{2(k+1)\pi^{k-\frac{1}{2}}}}Vol(\mathbb{H}^{2k+1})(k^2+m^2)^{k+\frac{1}{2}}\sum_{l=0}^{k-1}\frac{(-1)^l}{\Gamma(k-l+\frac{3}{2})(k^2+m^2)^l}a_{k,l}
\label{sec51 : eq5}
\feqr 
where the t-integral was performed by analytic continuation of the Gamma function to negative arguments. 
\par In general the integral in \rf{sec51 : eq2b} can fail to provide a finite result due to IR ($t\rightarrow \infty$) and UV ($t\rightarrow 0^+$) divergences. The first class of divergences cannot occur since the operator $\hat{D}_{\phi}$ is semi-bounded from below by $\lambda_0=(n-1)^2/4+m^2$ and the Green's function of $\hat{\Delta}_{\phi}$ is exponentially decreasing at infinity. As a consequence, a negative constant sectional curvature provides a natural infrared cut-off. The ultra violet divergency of $W^{(1-loop)}_{2k+1, div.}(\Lambda)$ can be found by introducing a lower cut off at $\epsilon=\Lambda^{-2}$, performing the partial integration and then taking the limit $\Lambda\rightarrow \infty$. In particular, for $\mathbb{H}^3$, $k=1$, and we would have
\beqr
W^{(1-loop)}_{\mathbb{H}^3}(\Lambda)
&=&\frac{Vol(\mathbb{H}^{3})}{2(4\pi)^{\frac{3}{2}}}b^{3}a_{1,0}\int_{\left(\frac{b}{\Lambda}\right)^{2}}^{\infty}u^{-\frac{5}{2}}e^{-u}du \nonumber \\
&=&W^{(1-loop)}_{\mathbb{H}^3, div.}(\Lambda)+W^{(1-loop)}_{\mathbb{H}^3, reg.}(\Lambda), \quad b^2=k^2+m^2
\label{sec51 : eq5a}
\feqr
where
\beqr
W^{(1-loop)}_{\mathbb{H}^3, div.}(\Lambda)&=&\frac{Vol(\mathbb{H}^{3})}{3(4\pi)^{\frac{3}{2}}}(m^2+1)^{\frac{3}{2}} a_{1,0}\left(-2\frac{\Lambda}{b}+\left(\frac{\Lambda}{b}\right)^3\right)e^{-\left(\frac{b}{\Lambda}\right)^2} \quad and
\label{sec51 : eq5b} \\
W^{(1-loop)}_{\mathbb{H}^3, reg.}(\Lambda)&=&\frac{Vol(\mathbb{H}^{3})}{12\pi}(m^2+1)^{\frac{3}{2}} a_{1,0}\, \textrm{Erf}\left(\frac{b}{\Lambda}\right)
\label{sec51 : eq5c}
\feqr
and $\textrm{Erf}(x)$ is the error function. Taking the $\Lambda\rightarrow 0^+$ limit of \rf{sec51 : eq5c},
\beqr
\lim_{\Lambda\rightarrow 0^+}W^{(1-loop)}_{\mathbb{H}^3, reg.}(\Lambda)=\frac{Vol(\mathbb{H}^{3})}{12\pi}(m^2+1)^{\frac{3}{2}}
\label{sec51 : eq5d}
\feqr
we recover the result of \cite{GM}. In the general case we find
\beqr
W^{(1-loop)}_{\mathbb{H}^{2k+1}}(\Lambda)&=&\frac{Vol(\mathbb{H}^{2k+1})}{2(4\pi)^{k+\frac{1}{2}}}b^{2k+1}\sum_{l=0}^{k-1}\frac{a_{k,l}}{b^{2l}}\int_{\left(\frac{b}{\Lambda}\right)^{2}}^{\infty}u^{l-k-\frac{3}{2}}e^{-u}du
\label{sec51 : eq5e}
\feqr
where the ultra-violet divergency and the regular part of the integral are both computed using the formula
\beqr
\int_{\left(\frac{b}{\Lambda}\right)^{2}}^{\infty}u^{-\left(s+\frac{1}{2}\right)}e^{-u}du &=& \frac{e^{-\left(\frac{b}{\Lambda}\right)^2}}{(2s-1)!!}\sum_{r=1}^{s-1}2^r(-1)^{r-1}(2s-2r-1)!!\left(\frac{\Lambda}{b}\right)^{2(s-r)+1} \nonumber \\
&+& \frac{(-1)^{s-1}2^s}{(2s-1)!!}\left[\left(\frac{\Lambda}{b}\right)e^{-\left(\frac{b}{\Lambda}\right)^2}-\int_{\left(\frac{b}{\Lambda}\right)^{2}}^{\infty}u^{-\frac{1}{2}}e^{-u}du\right]
\label{sec51 : eq5f}
\feqr
where $s=k-l+1$ is a positive integer.
\par In the even dimensional case, $n=2(k+1)$, the asymptotic behaviour of the heat kernel at coincident points is more involved. For $k=0$ we obtain (see Appendix C for details) 
\beqr
K_{2}(0,t)&=&\sqrt{\frac{4t}{\pi}}\frac{1}{4\pi t}e^{-(m^2+\frac{1}{4})t}\int_0^{\infty}\frac{u e^{-u^2}}{\sinh(\sqrt{t}u)} du \nonumber \\
&=&\frac{1}{4\pi t}e^{-(m^2+\frac{1}{4})t}\left(1+ \sum_{l=1}^{\infty} (2^{1-2l}-1)B_{2l}\frac{t^l}{l!}\right).
\label{sec51 : eq6}
\feqr 
In \rf{sec51 : eq6} $B_{2l}$ are the Bernoulli numbers and this expression is identical to the one derived in \cite{CA}. The ultra violet divergency in the one-loop partition function is given by
\beqr
W^{(1-loop)}_{\mathbb{H}^2, div.}(\Lambda)=\frac{Vol(\mathbb{H}^{2})}{4\pi}\left[\left(\Lambda^2-2b^2 \, ln\left(\frac{\Lambda}{b}\right)\right)e^{-\left(\frac{\Lambda}{b}\right)^2}+(m^2+\frac{1}{4})\gamma\right], \quad b^2=m^2+\frac{1}{4}
\label{sec51 : eq7}
\feqr
while the finite part is found to be 
\beqr
W^{(1-loop)}_{\mathbb{H}^2, reg.}=\frac{Vol(\mathbb{H}^{2})}{4\pi}(m^2+\frac{1}{4})\sum_{l=1}^{\infty}(2^{1-2l}-1)B_{2l}\frac{1}{l!(m^2+\frac{1}{4})^l}\Gamma(l-1)
\label{sec51 : eq7a}
\feqr
where $\gamma$ is the Euler's constant \rf{sec0 eq33}. In \rf{sec51 : eq7a} the series converges provided the mass for big $l$ behaves like $cl^2$ with the constant $c$ taking values greater than $1/\pi e$ as one may check using the asymptotic behaviour of Bernoulli's numbers $|B_{2l}|\approx 4\sqrt{\pi l}\,(l/2\pi)^{2l}$.
\par In the $\mathbb{H}^{4}$ case we use the recurrence relation \cite{DAV}
\beqr
K_{n+1}(r,t)=\sqrt{2}e^{\frac{1}{4}(2n+1)t}\int_r^{\infty}\frac{K_{n+2}(r,t)}{\sqrt{\cosh s-\cosh r}}\sinh s \, ds
\label{sec51 : eq7b}
\feqr
which connects the heat kernels in even and odd dimensions. Setting $n=3$, in \rf{sec51 : eq7b}, the integrand involves $K_5$ and its Maclaurin expansion around $t=0$ provides the asymptotic expansion
\beqr
K_{4}(0,t)=\frac{1}{(4\pi t)^2}e^{-\left(\frac{9}{4}+m^2\right)t}\!\!\left(\!\!1+\frac{t}{4}+\sum_{l=2}^{\infty}\frac{t^l}{l!}\left[(l-1)(2^{1-2l}-1)B_{2l}+\frac{l}{4}(2^{3-2l}-1)B_{2(l-1)}\right]\!\right)\!\!.
\label{sec51 : eq7c}
\feqr
Substituting \rf{sec51 : eq7c} back into \rf{sec51 : eq2b} we derive the one-loop UV divergency.
\par The simplest prescription for defining a finite one-loop partition function is to subtract from $K_n(x,x;t)$ the divergences proportional to $a_{k,l}$. It is easily  checked that the coefficient $\tilde{b}_{1,2k+1}=-k^2+a_{k,1}$ (or $\tilde{b}_{1,2(k+1)}=-k(k+1)-1/4+a_{2(k+1),1}$) of the linear term in $t$ equals to $-R/6$ where $R$ is the Ricci scalar curvature. On the other hand the coefficient $\tilde{b}_{2,2k+1}/2=k^4/2!-k^2a_{k,1}+a_{k,2}$ (or $\tilde{b}_{2,2(k+1)}/2=(k(k+1)+1/4)^2/2!-(k(k+1)+1/4)a_{2(k+1),1}+a_{2(k+1),2}$) of $t^2$ depends on a linear combination of $R_{\mu\nu \alpha \beta}^2, R_{\mu\nu}^2$ and $R^2$ (see \rf{sec51 : eq8c}). The following proposition provides an analytic result for the first three heat kernel coefficients in the scalar field case utilizing De Witt's iterative procedure \cite{BW}.
\begin{proposition}
The coincidence limits of the heat kernel coefficients 
\begin{displaymath}
\tilde{b}_{l,n}(0)\equiv \lim_{r\rightarrow 0^+}b_{0,n}(r)\equiv [b_{0,n}(r)],\,\, l=0,1,2, \,\, n=\textrm{dim}(\mathbb{H}^n)
\end{displaymath} 
are given by
\beqr
\tilde{b}_{0,n}&=&1 
\label{sec51 : eq8a} \\
\tilde{b}_{1,n}&=&-\frac{R}{6} 
\label{sec51 : eq8b} \\
\tilde{b}_{2,n}&=& \frac{n^4}{36}-\frac{n^3}{15}+\frac{13n^2}{180}-\frac{n}{30} \non \\
&=&\frac{1}{90}(R_{\mu\nu \alpha \beta}R^{\mu\nu \alpha \beta}-R_{\mu\nu}R^{\mu\nu})+\frac{R^2}{36}.
\label{sec51 : eq8c} \\
\feqr
\end{proposition} 
\textbf{Proof}\\
For the study of the asymptotic behaviour of $K_n$ in the $r\rightarrow 0^+$ limit a W.K.B. expansion suffices,    
\beqr
K_n(x,x',t)=\frac{1}{(4\pi t)^{\frac{n}{2}}}e^{-m^2 t}e^{-\frac{\sigma(x,x')}{2t}}\mathcal{P}(x,x')D^{\frac{1}{2}}(x,x')\Omega(x,x',t)
\label{sec51 : eq9}
\feqr
where $\sigma(x,x')=d^2_{\mathbb{H}^n}/2$ is the Synge's world function, $\mathcal{P}(x,x')$ is identically one in the neighbourhood of $x'$ but vanishes outside a normal neighbourhood of $x'$, 
\begin{displaymath}
D(x,x')=g^{-\frac{1}{2}}(x)\textrm{Det}(-\frac{\partial^2 \sigma(x,x')}{\partial_{\mu}\partial_{\nu'}})g^{-\frac{1}{2}}(x')
\end{displaymath} 
is the Van Vleck-Morette determinant which represents the density of geodesics. $\Omega$ is a function which satisfies the initial condition $\Omega(x,x',0)=1$ and is going to be determined. Substituting \rf{sec51 : eq9} into the heat equation and using the identities (see Appendix C for their proof)
\beqr
\sigma&=& \frac{1}{2}\sigma^{\mu}\sigma_{\mu}=\frac{1}{2}\sigma^{\mu'}\sigma_{\mu'}, \quad \sigma_{\mu}=\partial_{\mu}\sigma
\label{sec51 : eq9a}\\
n&=& D^{-1}\nabla_{\mu}\left(D\sigma^{\mu}\right)
\label{sec51 : eq9b} \\
D(r)&=& \left(\frac{r}{\sinh r}\right)^{n-1}
\label{sec51 : eq9c}
\feqr
we end up with the equation
\beqr
\left(\partial_t +\frac{1}{t}\sigma_{\mu} \partial^{\mu}-\mathcal{\hat{N}}\right)\Omega(x,x',t)=0, \quad \textrm{where} \quad \mathcal{\hat{N}}=D^{-\frac{1}{2}}\Box D^{\frac{1}{2}}.
\label{sec51 : eq10}
\feqr 
The solution that vanishes as $t$ goes to zero may be expressed as a power series
\beqr
\Omega(x,x',t)=\sum_{l=0}^{\infty}\frac{(-1)^l}{l!}b_{l,n}(x,x')t^l
\label{sec51 : eq10a}
\feqr
and the coefficients are determined by the differential recursion relation
\beqr
\left(1+\frac{1}{l}\sigma_{\mu} \partial^{\mu}\right)b_{l,n}(x,x')=\mathcal{\hat{N}}b_{l-1,n}(x,x')
\label{sec51 : eq10b}
\feqr
or 
\beqr
\left(1+\frac{1}{l}r\partial_r\right)b_{l,n}(r)=\mathcal{\hat{N}}b_{l-1,n}(r)
\label{sec51 : eq10c}
\feqr
by using the geodesic distance $r$. The solution of \rf{sec51 : eq10c} is 
\beqr
b_{l,n}(r)=lr^{l-1}\int_0^r \tilde{r}^{l-1}\mathcal{\hat{N}}b_{l-1,n}(\tilde{r})d\tilde{r}.
\label{sec51 : eq10d}
\feqr
For $l=0$, \rf{sec51 : eq8a} is obvious, bearing in mind the initial condition $\Omega(r,t=0)=1$. For $l=1$, \rf{sec51 : eq10d} gives
\beqr
b_{1,n}(r)=\frac{(n-1)}{4}\left[(n-3)\left(\frac{1}{r^2}-\frac{\coth(r)}{r}\right)+n-1 \right]
\label{sec51 : eq11}
\feqr
which reproduces \rf{sec51 : eq8b} in the coincidence limit $r\rightarrow 0^+$. The third heat kernel coefficient is recovered in the same way. The explicit expression of $b_{2,n}(r)$ is omitted due to its lengthy appearance. This method becomes very cumbersome beyond $\tilde{b}_{4,n}$ and a refined nonrecursive procedure was used by \cite{AV}, \cite{IA} to calculate $\tilde{b}_{8,n}$. For a vector field $\vec{\phi}$, having $N_0$ real components, relations \rf{sec51 : eq8a}-\rf{sec51 : eq8c} can be generalised by multiplying them by $N_0$.    
\par It is worth noting that the function 
\beqr
\Phi(r)=-D^{-1/2}(r)\Delta D^{1/2}(r)=-\frac{(n-1)}{4r^2}\left[(n-3)\left(1-r^2\coth^2 r\right)-2r^2\right],
\label{sec51 : eq12}
\feqr
is strictly positive, right-continuous, decreasing for $n=2$ and increasing for $n>3$. In addition
\beqr
\underset{r\in (0,\infty)}{\textrm{sup}\Phi(r)}&=&\frac{1}{3}, \quad \underset{r\in (0,\infty)}{\textrm{inf}\Phi(r)}=\frac{1}{4}, \quad \textrm{for} \quad n=2 \label{sec51 : eq12a1} \\
\underset{r\in (0,\infty)}{\textrm{sup}\Phi(r)}&=&\underset{r\in (0,\infty)}{\textrm{inf}\Phi(r)}=1, \quad \textrm{for} \quad n=3 \label{sec51 : eq12a2} \\
\underset{r\in (0,\infty)}{\textrm{sup}\Phi(r)}&=&\frac{(n-1)^2}{4}, \quad \underset{r\in (0,\infty)}{\textrm{inf}\Phi(r)}=\frac{n(n-1)}{6}, \quad \textrm{for} \quad n>3 
\label{sec51 : eq12a3}
\feqr
and satisfies the weighted Poincar\'{e} inequality 
\beqr
\int_{\mathbb{H}^n}|\nabla f|^2 d\mu\geq \int_{\mathbb{H}^n} \Phi f^2 d\mu\geq \underset{r\in \mathcal{D}\subset \mathbb{H}^n}{\textrm{inf}\, \Phi(r)}\int_{\mathbb{H}^n} f^2 d\mu
\label{sec51 : eq13}
\feqr 
w.r.t. the measure $d\mu(r)=(\sinh r)^{n-1} dr$ and for all compactly supported, smooth and radial functions $f\in \mathcal{C}^{\infty}_0(\mathbb{H}^n)$. In $n=2$ dimensions the $\underset{r\in \mathcal{D}\subset \mathbb{H}^n}{\textrm{inf}\, \Phi(r)}\geq 1/4$, as \rf{sec51 : eq12a1} requires, but not in $n>3$. Recall that $(n-1)^2/4$ is the greatest lower bound of the spectrum of the Laplacian acting on $\mathcal{L}^2$ functions. This inequality is generalized to every complete Riemannian manifold under certain conditions \cite{LW}.   
\subsection{$U(1)$ vector field}
\label{sec52}
Our starting point is a massive $U(1)$ gauge field $A_{\mu}$ on $\mathbb{H}^n$, with a $\xi$ dependent gauge fixed action, 
\beqr
S_E(A)&=&\int_{\mathbb{H}^n} \left(\frac{1}{4} F^{\mu \nu}F_{\mu \nu}+\frac{\xi}{2}(\nabla^{\mu}A_{\mu})^2 + \frac{1}{2}m^2 A_{\mu}A^{\mu}\right)\sqrt{g} d^n x \nonumber \\
&=& \frac{1}{2}\int_{\mathbb{H}^n}A_{\nu}\left(-\nabla^{\rho}\nabla_{\rho}A^{\nu}+\nabla_{\rho}\nabla^{\nu}A^{\rho}-\xi \nabla^{\nu}\nabla_{\rho}A^{\rho}+m^2 A^{\nu}\right)\sqrt{g} d^n x  \nonumber \\
&=&\frac{1}{2}\int_{\mathbb{H}^n}A^{\mu}\left(-g_{\mu \nu}\nabla^{\rho}\nabla_{\rho}-R_{\mu\nu}+(1-\xi)\nabla_{\mu}\nabla_{\nu}+g_{\mu \nu}m^2\right)A^{\nu}\sqrt{g} d^n x \nonumber \\
&=& \frac{1}{2}\int_{\mathbb{H}^n} A^{\mu}\hat{D}_{F,\mu\nu} A^{\nu}\sqrt{g} d^n x, \quad \hat{D}_{F,\mu\nu}=g_{\mu \nu}(-\Delta+m^2) - R_{\mu \nu}+(1-\xi)\nabla_{\mu}\nabla_{\nu}
\label{sec52 : eq1}
\feqr
where $F_{\mu \nu}=\nabla_{\mu}A_{\nu}-\nabla_{\nu}A_{\mu}$ is the field strength. We use the following conventions: $\nabla^{\mu}A_{\mu}=g^{\mu \nu}(\partial_{\nu}A_{\mu}-\Gamma_{\nu \mu}^{\rho}A_{\rho})$ and the components of the Ricci tensor are given by $R^{\lambda}_{\,\,\,\mu \lambda \nu}=\partial_{[\nu}\Gamma^{\lambda}_{\lambda]\mu}+\Gamma^{\sigma}_{\mu[\lambda} \Gamma^{\lambda}_{\nu]\sigma}$. The last definition leads to the identity $[\nabla_{\mu},\nabla_{\nu}]A^{\nu}=R_{\mu\nu}A^{\nu}$\footnote{In this notation the components of the Ricci tensor for the hyperbolic space are  positive.} with $R_{\mu \nu}=(n-1)g_{\mu \nu}$ for the hyperbolic space. The Green's function satisfies the equation
\beqr
\left(-\delta_{\mu}^{\nu}\nabla^{\rho}\nabla_{\rho}-R_{\mu}^{\nu}+(1-\xi)\nabla_{\mu}\nabla^{\nu}+\delta_{\mu}^{\nu}m^2\right)G_{\nu\nu'}(x,x')=\frac{g_{\mu\nu'}(x)}{\sqrt{g}}\delta^{(n)}(x-x').
\label{sec52 : eq1a}
\feqr
In the path integral approach of Quantum Field Theory one has to add the contribution of the Fadeev-Popov ghosts which in the Feynman gauge ($\xi=1$) is described by the action
\beqr
S_{ghost}=\int_{\mathbb{H}^n} \partial^{\mu}c \, \partial_{\mu}b\sqrt{g} d^n x =\int_{\mathbb{H}^n}c\hat{D}_{gh} b \sqrt{g} d^n x, \quad \hat{D}_{gh}=-\nabla^{\mu}\partial_{\mu}.
\label{sec52 : eq2}
\feqr
with $b$ and $c$ be real anti-commuting scalar fields. In the previous derivation of the actions \rf{sec52 : eq1} and \rf{sec52 : eq2} we again utilized Green's formula considering that $\textrm{suppA}$ is compact and either $\textrm{suppc}$ or $\textrm{supp$\nabla$b}$ is compact.
\par The total massless action, in the Feynman gauge, is invariant under the B.R.S.T. symmetry 
\beqr
\delta_{\theta}A_{\mu}=\theta \partial_{\mu}b, \quad \delta_{\theta}c=-\theta \nabla_{\mu}A^{\mu} \quad \delta_{\theta}b=0, \quad \theta^2=0
\label{sec52 : eq2b}
\feqr
which is parametrized by the infinitesimal anticommuting constant $\theta$ ($\{c,\theta\}=\{b,\theta\}=0$). 
\par The following bi-tensor identities for the derivatives of $u$ (see Appendix \rf{apA : eq4b}) are very crucial for solving the $U(1)$ heat equation
\beqr
\Box \left(\partial_{\nu}\partial_{\nu'}u\right)&=&\partial_{\nu}\partial_{\nu'}u, \quad \textrm{where} \quad \Box=g^{\mu \lambda}\nabla_{\mu}\nabla_{\lambda}\nonumber \\
\left(\nabla_{\mu}u\right) \nabla^{\mu}\left(\partial_{\nu}u\partial_{\nu'}u\right)&=& 4(1+u)\partial_{\nu}u\partial_{\nu'}u\nonumber \\
\Box\left(\partial_{\nu}u\partial_{\nu'}u\right)&=&(n+1)\partial_{\nu}u\partial_{\nu'}u+2(u+1)\partial_{\nu}\partial_{\nu'}u \nonumber \\
\left(\Box +n-1\right)\partial_{\nu}\partial_{\nu'}Q(t,u)&=&\partial_{\nu}\partial_{\nu'}\left(\Box Q(t,u)\right).
\label{sec52 : eq2ba}
\feqr 
The $U(1)$ kernel $K_{\mu \nu'}$ is a $(1,1)$ bi-tensor \cite{AJ} which can be written in the form (see Appendix \rf{apC : eq3})
\beqr
K_{\mu \nu'}(x,x';t)=F(t,u)\partial_{\mu}\partial_{\nu'}u+\partial_{\mu}\partial_{\nu'}Q(t,u)
\label{sec52 : eq2c}
\feqr
and solves the initial-boundary valued problem
\beqr
\partial_{t}K_{\mu \nu'}(x,x';t)&=&-\hat{D}_{F,\mu}^{\,\,\,\,\,\,\nu}K_{\nu \nu'}(x,x';t) \nonumber \\
K_{\mu \nu'}(x,x';0)&=& g_{\mu \nu'}(x)\delta^{(n)}(x,x').
\label{sec52 : eq3}
\feqr 
Using \rf{sec52 : eq2b} and \rf{sec52 : eq2c} one finds that
\beqr
-\hat{D}_{F,\mu}^{\nu}K_{\nu \nu'}(x,x';t)&=&\left(\partial_{\mu}\partial_{\nu'}u\right)\left(\Box +n-2-m^2\right)F+2\left(F\partial_{\mu}\partial_{\nu'}u+F'\partial_{\mu}u\partial_{\nu'}u  \right) \nonumber \\
&+& \partial_{\mu}\partial_{\nu'}\left(\Box Q\right)-m^2 Q \partial_{\mu}u\partial_{\nu'}u. 
\label{sec52 : eq3a}
\feqr
Substituting \rf{sec52 : eq3a} back into \rf{sec52 : eq3} we obtain the following partial differential equations
\beqr
\partial_t F(t,u)&=& (\Box_{\mathbb{H}^n}+(n-2)-m^2)F(t,u)
\label{sec52 : eq4a} \\
F(0,u)&=& -\delta^{(n)}(x,x') \label{sec52 : eq4b} \\
\partial_t Q(t,u)&=& \Box_{\mathbb{H}^n}Q(t,u)-m^2 Q(t,u)-2\int_{u}^{\infty}F(t,v)dv
\label{sec52 : eq4c} \\
\partial_u Q(0,u)&=& \partial_u^2 Q(0,u)=0 \label{sec52 : eq4d} 
\feqr
where $\Box_{\mathbb{H}^n}=u(u+2)\partial_u^2+n(u+1)\partial_u$. The solution of \rf{sec52 : eq4a} subjected to condition \rf{sec52 : eq4b} is 
\beqr
F_{2k+1}(t,r)=\frac{(-1)^{k+1}}{(2\pi)^k}\frac{e^{-m^2t}}{\sqrt{4\pi t}}\left(\frac{1}{\sinh r}\partial_r\right)^k e^{-\frac{r^2}{4t}}.
\label{sec52 : eq5}
\feqr
This is justified by noticing that \rf{sec3 : eq3a} receives now an extra contribution $-nf$ and, with the help of \rf{sec3 : eq3b} which remains unchanged, it leads to a time independent solution $f$, in the massless case. Therefore $\hat{B}(r,t)\propto \frac{1}{\sinh r} \frac{\partial}{\partial r}$.  
Fourier transforming equation \rf{sec52 : eq4d} we obtain, for $n=2k+1$, the kernel equation
\beqr
\partial_t \tilde{Q}(t,\lambda)=-(\lambda^2+\rho^2+m^2)\tilde{Q}(t,\lambda)+\frac{(-1)^{\rho-1}}{\pi(2\pi)^{\rho-1}\sqrt{4\pi t}}\tilde{F}_{2\rho-1}(t,\lambda).
\label{sec52 : eq6}
\feqr
The solution of \rf{sec52 : eq6} turns out to be
\beqr
\tilde{Q}(t,\lambda)&=&\frac{1}{\lambda^2+\rho^2+m^2}e^{-(\lambda^2+\rho^2+m^2)t} \nonumber \\
&+&\frac{2(-1)^{\rho-1}}{\lambda [(1+\rho^2+m^2)^2+4\lambda^2]}\left[(1+\rho^2+m^2)\sin(2t\lambda)-2\lambda\cos(2t\lambda)\right]e^{-(\lambda^2-1)t} \nonumber \\
&=& e^{-(\lambda^2+\rho^2+m^2)t} \Biggl[ \frac{1}{\lambda^2+\rho^2+m^2}+\frac{4(-1)^{\rho}}{[(1+\rho^2+m^2)^2+4\lambda^2]} \nonumber \\
&+& \frac{2(-1)^{\rho-1}t}{\lambda}\int_0^1 e^{(1+\rho^2+m^2)t\xi}\sin(2t\lambda \xi)d\xi\Biggr]
\label{sec52 : eq7}
\feqr
and the inverse Fourier transform is therefore given by
\beqr
Q(t,r)&=&\frac{1}{2^{\rho}\pi^{\rho+1}\Gamma(\rho)(\sinh r)^{2\rho-1}}\sum_{l_1=0}^{\rho-1}\left(\begin{array}{c} \rho-1 \\ l_1\end{array}\right)(-1)^{l_1} (\cosh r)^{l_1} \nonumber \\
&\times&\int_0^{\infty}\tilde{Q}(t,\lambda)\left(\int_{0}^{r}\cos(s\lambda)(\cosh s)^{\rho-1-l_1} ds\right)\prod_{l_2=1}^{\rho}\left[(\rho-l_2)^2+\lambda^2 \right]d\lambda.
\label{sec52 : eq8}
\feqr
The trace of the heat kernel for a $U(1)$ gauge field is
\beqr
Tr_{U(1)}^{\mathbb{H}^{2k+1}} K&=& \lim_{x'\rightarrow x}\left(g^{\mu \nu'}(x,x')K_{\mu \nu'}(t,x,x')\right) \nonumber \\
&=&-(2k+1)\lim_{r\rightarrow 0^+}\left(F(t,r)+\frac{1}{\sinh r}\partial_r G(t,r) \right)
\label{sec52 : eq9}
\feqr
since 
\beqr
&&[g^{\mu \nu'}(x,x')]=g^{\mu \nu}\label{sec52 : eq91}\\
&& [g^{\mu \nu'}\partial_{\mu}\partial_{\nu'}u]=-\textrm{dim}(\mathbb{H}^n)=-n \quad \textrm{and} 
\label{sec52 : eq10} \\
&& [g^{\mu \nu'}\partial_{\mu}u\,\partial_{\nu'}u]=0
\feqr
as one may prove by a straightforward calculation.
\par We investigate the massive-massless cases in $n=3$ dimensions. Using \rf{sec52 : eq5} we have
\beqr
\lim_{r\rightarrow 0^+}F_3(t,r)=-\frac{e^{-m^2 t}}{(4\pi t)^{\frac{3}{2}}}\lim_{r\rightarrow 0^+}\left(\frac{r}{\sinh r}e^{-\frac{r^2}{4t}}\right)=-\frac{e^{-m^2t}}{(4\pi t)^{\frac{3}{2}}}.
\label{sec52 : eq11}
\feqr
On the other hand from \rf{sec52 : eq8} we obtain
\beqr
&&\!\!\!\!\!Q(t,r)= \frac{e^{-t(1+m^2)}}{2|\mathbb{S}^2|\sinh r}\Biggr[2\left[e^{t(1+\frac{m^2}{2})^2}\sinh(r(1+\frac{m^2}{2}))-e^{t(1+m^2)}\sinh(r\sqrt{1+m^2})\right]
\nonumber \\
&+&\!\!\!\!\!e^{t(1+\frac{m^2}{2})^2}e^{-r(1+\frac{m^2}{2})} \textrm{Erf}\left(\!\!\sqrt{t}(1+\frac{m^2}{2})\!-\frac{r}{2\sqrt{t}}\!\right)\!-e^{t(1+m^2)} e^{-r\sqrt{1+m^2}}\textrm{Erf}\left(\!\!\sqrt{t(1+m^2)}-\frac{r}{2\sqrt{t}}\!\right) \nonumber \\
&+&\!\!\!\!\! e^{t(1+m^2)} e^{r\sqrt{1+m^2}}\textrm{Erf}\left(\!\!\sqrt{t(1+m^2)}\!+\!\frac{r}{2\sqrt{t}}\!\right)\!-e^{t(1+\frac{m^2}{2})^2}e^{r(1+\frac{m^2}{2})} \textrm{Erf}\left(\!\!\sqrt{t}(1+\frac{m^2}{2})\!+\!\frac{r}{2\sqrt{t}}\!\right)\nonumber \\
&+&\!\!\!\!\!\sqrt{\frac{t}{\pi}}\frac{2e^{-\frac{r^2}{4t}}}{|\mathbb{S}^2|\sinh r}\int_0^1 e^{-tm^2(1-\xi)}e^{-t(1-\xi)^2}\sinh(r\xi)d\xi.
\label{sec52 : eq12}
\feqr
The partial derivative of \rf{sec52 : eq12} and the integral of the last term can be performed exactly. We only present now the massless case for simplicity, which gives
\beqr
\lim_{m\rightarrow 0}\lim_{r\rightarrow 0}\left(\frac{1}{\sinh r}\partial_r Q(t,r) \right)=-\frac{(4t+e^{-t}-1)}{3(4\pi t)^{\frac{3}{2}}}.
\label{sec52 : eq13}
\feqr  
Finally the trace of the heat kernel for a massless vector field in $\mathbb{H}^3$ is
\beqr
Tr K_{U(1)}^{\mathbb{H}^{3}}=\frac{(2+4t+e^{-t})}{(4\pi t)^{\frac{3}{2}}}.
\label{sec52 : eq14}
\feqr
\begin{proposition}
The coincidence limits of the first three heat kernel coefficients, in the presence of a $U(1)$ gauge field, are given by
\beqr
\tilde{b}_{0,\mu\nu'}&=& g_{\mu\nu'}(x,x) 
\label{sec52 : eq15a} \\
\tilde{b}_{1,\mu\nu'}&=& \frac{1}{n}\left(1-\frac{n}{6}\right)R \,g_{\mu\nu'}
\label{sec52 : eq15b} \\
\tilde{b}_{2,\mu\nu'}&=& \left(\frac{n^4}{72}-\frac{n^3}{5}+\frac{313 n^2}{360}-\frac{27 n}{20}+\frac{2}{3}\right)g_{\mu\nu'}
\label{sec52 : eq15c} 
\feqr
where $g_{\mu \nu'}(x,x')$ is the parallel displacement operator of the field along the geodesic from the point $x'$ to the point $x$. The $U(1)$ trace of  $\tilde{b}_{2,\mu \nu'}(x,x)$ is then given by 
\beqr
\textrm{Tr}_{U(1)}^{\mathbb{H}^{n}} b_2=\frac{1}{360}\left((2n-30)R_{\mu \nu \alpha \beta}R^{\mu \nu \alpha \beta}+(180-2n)R_{\mu\nu}R^{\mu\nu}+(5n-60)R^2\right).
\label{sec52 : eq15d}
\feqr 
\end{proposition}
\textbf{Proof}\\
Substituting the W.K.B. expansion of the $U(1)$ heat kernel 
\beqr
K_{\mu \nu'}(x,x',t)&=&\frac{1}{(4\pi t)^{\frac{n}{2}}}e^{-m^2 t}e^{-\frac{\sigma(x,x')}{2t}}D^{\frac{1}{2}}(x,x')\Omega_{\mu \nu'}(x,x',t) \quad \textrm{with} \nonumber \\
\lim_{t\rightarrow 0^+}K_{\mu \nu'}(x,x',t)&=&g_{\mu\nu'}(x,x)\delta^{(n)}(x-x')
\label{sec52 : eq16}
\feqr
back into \rf{sec52 : eq3} and using the power series expansion 
\beqr
\Omega_{\mu \nu'}(x,x',t)=\sum_{l=0}^{\infty} b_{l,\mu \nu'}(x,x')t^l
\label{sec52 : eq17}
\feqr
we obtain the following recurrent differential equation
\beqr
&&\left(1+\frac{1}{l}\sigma_{\lambda} \nabla^{\lambda}\right)b_{l,\mu \nu'}(x,x')=\frac{1}{l}\left(\hat{\mathcal{N}}+n-1\right)b_{l-1,\mu \nu'}(x,x'), \quad l\geq1 \quad \textrm{and} 
\label{sec52 : eq18a} \\
&&\sigma_{\lambda} b_{0,\mu \nu';}^{\quad\quad\lambda}(x,x')=0
\label{sec52 : eq18b}
\feqr
where the semicolon denotes covariant derivative in \rf{sec52 : eq18b}. The solution of \rf{sec52 : eq18b} is $b_{0,\mu \nu'}(x,x')=g_{\mu \nu'}(x,x')$ which in the $x'\rightarrow x$ limit becomes the unit matrix. The recurrent partial differential equation \rf{sec52 : eq18a} at the coincidence limit is rewritten as
\beqr
\tilde{b}_{2,\mu\nu'}&=&\frac{1}{2}\left[\hat{\mathcal{N}}b_{1,\mu\nu'}(x,x')\right]
+\frac{(n-1)}{2}\tilde{b}_{1,\mu\nu'}.\label{sec52 : eq18c}
\feqr
Using the identity
\beqr
\Box b_{1,\mu \nu'}(x,x')=\Box \left(\hat{\mathcal{N}}b_{0,\mu \nu'}(x,x')+(n-1)b_{0,\mu \nu'}(x,x')-d\eta^{\lambda} b_{1,\mu \nu';\lambda}(x,x')\right)
\label{sec52 : eq19}
\feqr 
and the coincidence limits
\beqr
\left[\Box \mathcal{\hat{N}}\right]&=&-\frac{n}{30}(n-3)(n-1)\label{sec52 : eq19a} \\
\left[\mathcal{\hat{N}}(\Box g_{\mu\nu'}(x,x')\right]&=&-(n-1) g_{\mu\nu'}\label{sec52 : eq19b} \\
\left[D^{-\frac{1}{2}}(\Box D^{-\frac{1}{2}})b_{1,\mu \nu'}(x,x')\right]&=&-\frac{n(n-1)^2}{6}\left(1-\frac{n}{6}\right)g_{\mu\nu'}\label{sec52 : eq19c} 
\feqr
we find that 
\beqr
\left[\hat{\mathcal{N}}b_{1,\mu\nu'}(x,x')\right]=\left(\frac{n^4}{36}-\frac{7n^3}{30}-\frac{73n^2}{180}+\frac{8n}{15}-\frac{1}{3}\right)g_{\mu\nu'}
\label{sec52 : eq19d}
\feqr
which combined with \rf{sec52 : eq18c} the second heat kernel coefficient \rf{sec52 : eq15c} is recovered. The $U(1)$ trace of the heat kernel, in the massless case, is identical to the one predicted by the Fourier method in $n=3$ dimensions.
\par The general solution of the recurrent equation \rf{sec52 : eq18b}, for arbitrary value of $l\geq 1$, can be written formally as
\beqr
b_{l,\mu \nu'}(x,x')=\frac{1}{l!}\left(1+\frac{1}{l}\hat{A}\right)^{-1}\hat{F}\left(1+\frac{1}{l-1}\hat{A}\right)^{-1}\hat{F}\cdots \left(1+\hat{A}\right)^{-1}b_{0,\mu \nu'}(x,x')
\label{sec52 : eq21}
\feqr
where $\hat{A}=\sigma_{\lambda}\nabla^{\lambda}$ and $\hat{F}=\hat{N}+n-1$. Adding the contribution of the ghosts fields the $U(1)$ traces in the massless case become
\beqr
\textrm{Tr}_{U(1)}^{\mathbb{H}^{n}} b_{0,tot.}&=&n-2
\label{sec52 : eq22a} \\
\textrm{Tr}_{U(1)}^{\mathbb{H}^{n}} b_{1,tot.}&=&\frac{1}{6}(8-n)R
\label{sec52 : eq22b} \\
\textrm{Tr}_{U(1)}^{\mathbb{H}^{n}} b_{3,tot.}&=&\frac{1}{180}\left((n-17)R_{\mu \nu \alpha \beta}R^{\mu \nu \alpha \beta}+(92-n)R_{\mu\nu}R^{\mu\nu}\right)+\frac{1}{72}(n-14)R^2.
\label{sec52 : eq22c} \\
\feqr 
\section{Conclusions}
\label{conclus}

In this paper we have calculated the one-loop partition function for a free, massive, real quantum scalar field and $U(1)$ gauge field, living on the hyperbolic space $\mathbb{H}^n(\mathbb{R})$. In the scalar case we have provided closed expressions for the regular and UV divergent parts of the partition function. The $U(1)$ vector case was solved by adapting a suitable ansatz and applying Fourier analysis on the corresponding heat equation. The regular parts of both  partition functions were justified by an independent W.K.B. approximation in any dimension. We have also proposed an alternative approach to derive recurrent relations of the fundamental solutions of the heat equation in different dimensions.           

\appendix
\section*{Appendices}
\section*{Appendix A}
\label{apA}
\renewcommand{\theequation}{A.\arabic{equation}}
\renewcommand{\theproposition}{A.\arabic{proposition}}
\setcounter{equation}{0}

\par The Hodge inner product of a $p-$form ($p\leq n-1$), taking values on a real vector bundle $E$, is defined generally by 
\beqr
\langle a_p,a_p\rangle=(-1)^s \int_{\mathcal{M}_n} a_p\wedge \star a_p=\frac{(-1)^s}{p!}\int_{\mathcal{M}_n}a_{i_1\cdots i_p}a^{i_1\cdots i_p}\sqrt{g}d^nx
\label{apA : eq1a} 
\feqr
where the index $s$ denotes the dimension of the maximal subspace on which $g$ is negative definite. For the real hyperbolic space $s=0$. 
\par There exist two distinct coordinate representations of the hyperbolic space that will be utilized throughout the present work.
\begin{description}
\item [\textit{The geodetic spherical polar representation.}] The equation of hyperboloid is satisfied by the transformations
\beqr
X^1 &=& \sinh r \cos \theta_1, \quad r\in(0,\infty), \,\, \theta_1\in[0,2\pi) \nonumber \\
X^2&=& \sinh r \sin \theta_1 \cos \theta_2  \nonumber \\
&\vdots& \nonumber \\
X^{n-1}&=& \sinh r \sin \theta_1 \sin \theta_2 \cdots \cos \theta_{n-1} \nonumber \\
X^n&=& \sinh r \sin \theta_1 \sin \theta_2 \cdots \sin \theta_{n-1}, \quad \theta_k\in[0,\pi], \,\, k=2,\cdots, n-1 \nonumber \\
X^{0}&=& \cosh r
\label{apA : eq1}
\feqr
and the space has the topology of $\mathbb{R}^+\times \mathbb{S}^{n-1}$. 
A direct calculation leads to the metric tensor and the volume element in the form  
\beqr
ds_{\mathbb{H}^n}^2&=&dr^2+\sinh^2 r \, \gamma_{ij} \,d\theta^i d\theta^j
\label{apA : eq2}\\
dV&=& (\sinh r)^{n-1}drd\Omega_{n-1}
\label{apA : eq21}
\feqr
where $\gamma_{ij}$ is the metric on $\mathbb{S}_{n-1}$ and $d\Omega_{n-1}$ its volume element. Note that the hyperbolic spherical polar coordinates are the same as the Euclidean ones with the substitution $\sinh r$ for $r$. The Laplace-Beltrami operator is given by
\beqr
\Delta_{\mathbb{H}^n}=\partial_r^2+(n-1)\coth r\, \partial_r +\frac{1}{\sinh^2 r}\Delta_{\mathbb{S}^{n-1}}.
\label{apA : eq2a}
\feqr
\item [\textit{The Poincar\'{e} half-space representation.}]
The second class of transformations that preserves the bilinear form is
\beqr
X^i&=& \frac{x_i}{x_n}, \quad i=1,\cdots, n-1 \nonumber \\
X^n+X^{0}&=& \frac{1}{x_n} \nonumber \\
X^n-X^{0}&=& \frac{1}{x_n}\sum_{i=1}^{n}x_i^2
\label{apA : eq3}
\feqr
The metric tensor  
\beqr
ds_{\mathbb{H}^n}^2=\frac{1}{x_n^2}\sum_{i=1}^{n}(dx_i)^2
\label{apA : eq4} 
\feqr
is manifestly conformally invariant and superior in some calculations as compared to \rf{apA : eq1}. The Laplace -Beltrami operator in this model is
 \beqr
\Delta_{\mathbb{H}^n}=x_n^2\Delta_{\mathbb{E}^n}-(n-2)x_n\partial_n
\label{apA : eq4a}
\feqr
where $\Delta_{\mathbb{E}^n}$ is the Euclidean Laplacian. 
\end{description}
\begin{proposition}
The geodesics of $\mathbb{H}^n$ are straight lines perpendicular to the hyperplane $x_n=0$ (the boundary $\partial \mathbb{H}^n$) and circles whose planes are perpendicular to $x_n=0$ with centers on the hyperplane.
\end{proposition} 
The proof is based on finding the solutions to the geodesics equation
\beqr
\frac{d^2 x^{\mu}}{d\tau^2}+\Gamma^{\mu}_{\nu \lambda}\frac{dx^{\nu}}{d\tau}\frac{dx^{\lambda}}{d\tau}=0
\label{apA : eq4ab}
\feqr 
where $\tau$ is an affine parameter and
\begin{displaymath}
\Gamma^{\mu}_{\nu \lambda}=\frac{1}{2} g^{\mu \sigma}\left(\partial_{\nu}g_{\lambda \sigma}+\partial_{\lambda}g_{\nu \sigma}-\partial_{\sigma}g_{\nu \lambda} \right), \quad g_{\mu\nu}=\frac{1}{x_n^2}\delta_{\mu \nu}
\end{displaymath}
are the components of the Levi-Civita connection (the well-known Christoffel symbols of the second kind). The Levi-Civita connection is an affine connection which is metric compatible and torsion free.  
Considering two points on $\mathbb{H}^n$ their geodesic distance is given by 
\beqr
d_{\mathbb{H}^n}(x,x')=\textrm{inf}\Biggl\{\int_0^1\sqrt{g_{\mu \nu}(x(\tau)) \dot{x}^{\mu}\dot{ x}^{\nu}} d\tau, x(\tau)\in C^1([0,1]),\, x(0)=x, \, x(1)=x'  \Biggr\}
\label{apA : eq4b0}
\feqr
which is found to be  
\beqr
\cosh d_{\mathbb{H}^n}(x,x')=1+\frac{1}{2x_nx'_n}d_{\mathbb{E}_n}^2(x,x')=1+u(x,x')
\label{apA : eq4b}
\feqr
and $u(x,x')$ is the \textit{chordal distance}. Using the half-space representation one can have the alternative expression for $u(x,x')$
\beqr
u(x,x')&=&\frac{1}{2}\eta_{AB}(X-X')^{A}(X-X')^{B}=\frac{\delta_{ij}(x-x')_i(x-x')_j}{2x_nx_n'}, \nonumber \\
 A,B&=&0,\cdots, n, \quad i,j=1,\cdots,n
\label{apA : eq4c}
\feqr
where the Minkowski metric $\eta_{AB}$ has signature $(-,+,\cdots,+)$ and Einstein's summation convention for each repeated pair of indices is adopted. 
\par In general a function of the hyperbolic distance satisfies the equation
\beqr
\Delta f(d_{\mathbb{H}^n})=(n-1)f'(d_{\mathbb{H}^n})\coth (d_{\mathbb{H}^n})+f''(d_{\mathbb{H}^n}).
\label{apA : eq4d}
\feqr
This can be proved by first showing the following relation 
\beqr
x_n^2\left(\Delta_{\mathbb{E}^{n-1}}+\partial_n^2\right)d_{\mathbb{H}^n}=1.
\label{apA : eq4e}
\feqr 
\section*{Appendix B}
\label{apB}
\renewcommand{\theequation}{B.\arabic{equation}}
\setcounter{equation}{0}
\par In the Fourier transform we are going to use a form of $\Phi_{\lambda}$ which is derived by changing $r$ into the variable $s=-\ln(\cosh r-\cos \theta \sinh r)$ and thus obtaining
\beqr
\Phi_{\lambda}(r)\!\!\!&=&\!\!\! \frac{(-1)^{\rho +1}2^{\rho-1}\Gamma\left(\rho +\frac{1}{2}\right)}{(\sinh r)^{2\rho -1}\sqrt{\pi}\Gamma(\rho)}\int_{-r}^{r}e^{is\lambda}(\cosh r -\cosh s)^{\rho-1} ds \nonumber \\
\!\!\!&=&\!\!\! \frac{2^{\rho}\Gamma\left(\rho +\frac{1}{2}\right)}{(\sinh r)^{2\rho -1}\sqrt{\pi}\Gamma(\rho)}\sum_{l_1=0}^{\rho-1}\!\left(\begin{array}{c} \rho-1 \\ l_1\end{array}\right)\!(-1)^{l_1} (\cosh r)^{l_1}\!\!\!\int_{0}^{r}\!\!\!\cos(s\lambda)(\cosh s)^{\rho-1-l_1} ds.
\label{apB : eq4c}
\feqr
Also the modulus of the Harish-Chandra $c$-function is given by
\beqr
|c(\lambda)|^2 &=&\frac{4^{2\rho-1}\Gamma^2\left(\rho+\frac{1}{2}\right)}{\pi}\left|\frac{\Gamma(i\lambda)}{\Gamma(\rho+i\lambda)}\right|^2 \nonumber \\
&=& \frac{4^{2\rho-1}\Gamma^2\left(\rho+\frac{1}{2}\right)}{\pi}\frac{1}{\prod_{l_2=1}^{\rho}\left[(\rho-l_2)^2+\lambda^2 \right]}
\label{apB : eq4d}
\feqr
where the gamma function identities, for $z\in\mathbb{C}$, 
\beqr
\overline{\Gamma(z)}&=& \Gamma(\overline{z}) \label{apB : eq4e1} \\
\Gamma(1+z)&=& z\Gamma(z)\label{apB : eq4e2} \\
\Gamma(1-z)\Gamma(z)&=&\frac{\pi}{\sin(\pi z)} \label{apB : eq4e3}.
\feqr
have been used.
\par The derivation of the one-loop effective actions \rf{sec51 : eq5a}, \rf{sec51 : eq7}, \rf{sec51 : eq7a}, on $\mathbb{H}^3, \mathbb{H}^2$, is based on $p_3, p_5$ which are obtained by applying the recurrence relation
\beqr
p_{2k+1}(r,t)=-\frac{1}{2\pi}e^{-(2k-1)t}\frac{1}{\sinh r}\frac{\partial}{\partial r}p_{2k-1}(r,t).
\label{apB : eq4b0}
\feqr
We thus have
\beqr
p_3(r,t)\!\!\!&=&\!\!\! \frac{1}{(4 \pi t)^{\frac{3}{2}}}\frac{r}{\sinh r}e^{-\left(\frac{r^2}{4t}+t\right)}
\label{apB : eq4b1} \\
p_5(r,t)\!\!\!&=&\!\!\! \frac{1}{(4 \pi t)^{\frac{5}{2}}}\left[\left(\frac{r}{\sinh r}\right)^2+g(r,t)\right]e^{-\left(\frac{r^2}{4t}+4t\right)}, \,\, g(r,t)=2t\frac{(r \cosh r- \sinh r)}{\sinh^3 r}
\label{apB : eq4b2} \\
p_7(r,t)\!\!\!&=&\!\!\! \frac{1}{(4 \pi t)^{\frac{7}{2}}}\left[\left(\frac{r}{\sinh r}\right)^3+\frac{3r}{\sinh r}g(r,t)-\frac{2t}{\sinh r}\frac{\partial}{\partial r}g(r,t)\right]e^{-\left(\frac{r^2}{4t}+9t\right)} \nonumber \\
&\textrm{where}& \frac{\partial}{\partial r}g(r,t)=\frac{2t}{\sinh^2 r}(r+3\coth r -3r\coth^2 r)
\label{apB : eq4b3}
\feqr
It is interesting to observe that all the heat kernels in $\mathbb{H}^{2k+1}$ are given by elementary functions while in even dimensions by integrals.
\par For $\mathbb{H}^3$ we have $a_{1,0}=1$, $\mathbb{H}^5$ $(a_{2,0},a_{2,1})=(1, 2/3)$ and $\mathbb{H}^7$ $(a_{3,0},a_{3,1},a_{3,2})=(1,2,16/15)$. Note that 
\beqr
a_{k+1,0}=\lim_{r\rightarrow 0^+}\left(\frac{r}{\sinh r}\right)^k=1, \, \forall k\in\mathbb{N}.
\label{apB : eq4e31}
\feqr
\par In the derivation of \rf{sec51 : eq6} we used the trigonometric identity
\beqr
\sinh(\frac{x}{2})=\sqrt{\frac{\cosh x -1}{2}} sign(x),
\label{apB : eq4e3a}
\feqr
the Taylor expansion of $1/\sinh x$ around $x=0$, namely
\beqr
\frac{1}{\sinh x}=\frac{1}{x}\left(1-\sum_{l=1}^{\infty}2(2^{2l-1}-1)B_{2l} \frac{x^{2l}}{(2l)!}\right)
\label{apB : eq4e3b}
\feqr
and
\beqr
\frac{2^l (2l-1)!!}{(2l)!}=\frac{1}{l!}.
\label{apB : eq4e3c}
\feqr
Another useful Maclaurin expansion, used in \rf{sec51 : eq7c}, is 
\beqr
\frac{1}{\cosh x}=\sum_{k=0}^{\infty}\frac{(-1)^k 2(1-2^{k+1})B_{k+1}}{(k+1)(2k)!}x^{2k}.
\label{apB : eq4e3d}
\feqr
\section*{Appendix C}
\label{apC}
\renewcommand{\theequation}{C.\arabic{equation}}
\setcounter{equation}{0}
The proof of \rf{sec51 : eq9c} is based on the following relation
\beqr
D(x,x')&=&\left(\frac{r}{\sinh r}\right)^n \textrm{Det} \mathcal{A}_{\mu \nu'}, \quad \textrm{where}
\label{apC : eq4f} \\
\mathcal{A}_{\mu \nu'}&=& \frac{1}{\sinh r}\left(\coth r -\frac{1}{r}\right)\partial_{\mu}u \partial_{\nu'}u+\partial_{\mu}\partial_{\nu'}u \nonumber \\
&=&\frac{1}{\sinh r}\left(\coth r -\frac{1}{r}\right)\left(\frac{(x-x')_{\mu}}{x'_n}-u\delta_{n\mu}\right) \left(\frac{(x'-x)_{\nu'}}{x_n}-u\delta_{n\nu'}\right)\nonumber \\
&+& \left(\delta_{\mu \nu'}+\frac{(x-x')_{\mu}}{x'_n}\delta_{n\nu'}+\frac{(x'-x)_{\nu'}}{x_n}\delta_{n\mu}-u\delta_{n\nu'}\delta_{n\mu} \right) \quad \textrm{and} \nonumber \\
\textrm{Det} \mathcal{A}_{\mu \nu'}&=&\frac{\sinh r}{r}.
\feqr
The ansatz \rf{sec52 : eq2c} we used for the $U(1)$ heat kernel is justified by the observation that any $(1,1)$ maximally symmetric bi-tensor can be expressed as the following combination
\beqr
K_{\mu \nu'}(x,x';t)=f_1(t,u)g_{\mu \nu'}(x,x')+f_2(t,u)\eta_{\mu}(x,x')\eta_{\nu'}(x,x')
\label{apC : eq1}
\feqr 
where $\eta_{\mu}(x,x')=\partial_{\mu}d_{\mathbb{H}^n}(x,x')$ and $\eta_{\nu'}(x,x')=\partial_{\nu'}d_{\mathbb{H}^n}(x,x')$ are the unit tangents to the chordal distance at $x$ and $x'$ respectively. Taking into account the relation 
\beqr
\nabla_{\mu}\eta_{\nu'}=-\frac{1}{\sinh d}(g_{\mu \nu'}+\eta_{\mu}\eta_{\nu'})
\label{apC : eq2}
\feqr
equation \rf{apC : eq1} can be rewritten as
\beqr
K_{\mu \nu'}(t,u)&=&A(t,u)\partial_{\mu}\partial_{\nu'}u+B(t,u) \, \partial_{\mu}u \partial_{\nu'} u \quad \textrm{where} \\
A(t,u)&=&\frac{1}{u+2}\left(f_1(t,u)+\frac{f_2(t,u)}{u}\right) \quad \textrm{and} \quad B(t,u)=-f_1(t,u). \nonumber
\label{apC : eq3}
\feqr
\par The derivation of \rf{sec52 : eq12} requires the following integrals 
\beqr
\int_0^1\!\!\! e^{(1+\rho^2+m^2)t\xi}\sin(2t\lambda \xi)d\xi\!\!\!\!=\!\!\!\!\frac{\left[2\lambda + e^{(1+\rho^2+m^2)t}\left[(1+\rho^2+m^2)\sin(2t\lambda)-2\lambda \cos(2t \lambda)\right]\right]}{t[(1+\rho^2+m^2)^2+4\lambda^2]}
\label{apC : eq4}
\feqr
and
\beqr
\int_0^{\infty}\!\!\!\!e^{-\beta x^2}\sin(\alpha x)\frac{x}{(\gamma^2+x^2)}dx \!\!\!\!&=&\!\!\!\!-\frac{\pi}{4}e^{\beta \gamma^2}\Bigr[2\sinh(\alpha \gamma)+e^{-\alpha \gamma} \textrm{Erf}\left(\!\!\gamma \sqrt{\beta}-\frac{\alpha}{2\sqrt{\beta}}\right) \nonumber \\
\!\!\!\!&-&\!\!\!\!e^{\alpha \gamma} \textrm{Erf}\left(\!\!\gamma \sqrt{\beta}+\frac{\alpha}{2\sqrt{\beta}}\right)\Bigl].
\label{apC : eq5}
\feqr
\textbf{Useful identities and proofs involving coincident limits}\\
Using the definition of the Synge's function $\sigma(x,x')=d^2_{\mathbb{H}^n}(x,x')/2$ \cite{JS} we can prove the following identities ($d\equiv d_{\mathbb{H}^n}$):
\beqr
\eta_{\mu'}&=&-g_{\mu'}^{\,\,\,\nu}\eta_{\nu}, \quad \eta_{\mu}\eta^{\mu}=\eta_{\mu'}\eta^{\mu'}=1
\label{apC : eq5a1} \\
\eta_{\mu;\nu}&=&A(d)(g_{\mu \nu}-\eta_{\mu}\eta_{\nu}), \quad \textrm{where} \quad A(d)=\coth d \label{apC : eq5a2} \\
\eta_{\mu';\nu}&=&C(d)(g_{\mu' \nu}+\eta_{\mu'}\eta_{\nu}), \quad \textrm{where} \quad C(d)=-\frac{1}{\sinh d} \label{apC : eq5a3} \\
\eta^{\nu}\eta_{\mu;\nu}&=&\eta^{\nu}\eta_{\mu';\nu}=0\label{apC : eq5a4} \\
\Box\eta_{\mu}&=&-(n-1)A^2\eta_{\mu}, \quad \Box\eta_{\mu'}=-(n-1)C^2\eta_{\mu'}\label{apC : eq5a5} 
\feqr
\beqr
g_{\mu' \nu;\lambda}&=&-(A+C)(g_{\nu\lambda}\eta_{\mu'}+g_{\mu'\lambda}\eta_{\nu})\label{apC : eq5a6} \\
\Box g_{\mu' \nu}&=&-(A+C)^2(g_{\mu' \nu}-(n-2)\eta_{\mu'}\eta_{\nu})\label{apC : eq5a7} \\
\eta^{\lambda}g_{\mu'\nu;\lambda}&=&0\label{apC : eq5a8} \\
\sigma_{\mu;\nu}&=& d A g_{\mu\nu}+(1-d A)\eta_{\mu}\eta_{\nu}, \,\, \sigma_{\mu}=d\eta_{\mu}\label{apC : eq5a9} \\
\sigma_{\mu';\nu}&=& d C g_{\mu'\nu}+(1+d C)\eta_{\mu'}\eta_{\nu}\label{apC : eq5a10}\\
\sigma_{\mu;\nu;\lambda}-\sigma_{\mu;\lambda;\nu}&=&R^{\rho}_{\,\,\,\mu\nu\lambda}\sigma_{\rho}\label{apC : eq5a11}
\feqr
Next we list some identities related to coincidence limits
\beqr
\left[\eta_{\mu}\right]&=&\left[\eta_{\mu'}\right]=\left[\sigma_{\mu}\right]=\left[\sigma_{\mu'}\right]=\left[D_{\mu}\right]=0 \quad \textrm{and} \quad \left[D\right]=1
\label{apC : eq5b1}\\
\left[\sigma_{\mu ; \nu}\right]&=&-\left[\sigma_{\mu'; \nu}\right]=-\left[g_{\mu'\nu}\right]=-g_{\mu\nu}\label{apC : eq5b2}\\
\left[D_{\mu;\nu}^{\frac{1}{2}}\right]&=&-\frac{1}{6}R_{\mu\nu}, \quad \left[\Box D^{\frac{1}{2}}\right]=-\frac{1}{6}R\label{apC : eq5b3}\\
\left[\Box \Box g_{\mu'\nu}\right]&=&-(n-1)g_{\mu'\nu}\label{apC : eq5b4}.
\feqr
We now establish the proof of relation \rf{sec51 : eq9b}.
Observe that
\beqr
D^{-1}\nabla_{\mu}(D \partial \sigma^{\mu})&=&\Box \sigma +d D^{-1}D', \quad \textrm{where} \quad D'\equiv \frac{dD(r)}{dr} \label{apC : eq7}\\
&=&n
\feqr
since $\Box \sigma=d(n-1) \coth  (d)+1$ by \rf{apA : eq4d}.


\begin{thebibliography} {00}

\bibitem{AJ} B. Allen and T. Jacobson, ``Vector two point functions in maximally symmetric spaces", \cmp{103}{1986}{669}.

\bibitem{AV} I. G. Avramidi, ``A covariant technique for the calculation of the one-loop effective action", \np{355}{1991}{712-754}, Erratum:ibid \np{509}{1998}{557-558}. 

\bibitem{IA} I. G. Avramidi, \textit{Heat Kernel and Quantum Gravity}, Lecture Notes in Physics Monographs, Springer-Verlag Berlin Heidelberg, 2000. 

\bibitem{B} A. A. Bytsenko, G. Cognola, L. Vanzo and S. Zerbini, ``Quantum Fields and Extended Objects in Space-Times with Constant Curvature Spatial Section", \pr{266}{1996}{1-126}.

\bibitem{OB} W. O. Bray, ``Aspects of Harmonic Analysis on Real Hyperbolic Space", in ``Fourier Analysis: Analytic and Geometric Aspects", (W. O. Bray, P. S. Milojevic and Caslav V. Stanojevi Eds.) Vol. 157, pp. 77-102, Lecture Notes in Pure and Applied Mathematics, Marcel Dekker, New York, 1994.

\bibitem{CA} R. Camporesi, ``Harmonic analysis and propagators on homogeneous spaces", \pr{196}{1990}{1}.

\bibitem{CH} I. Chavel, \textit{Riemannian Geometry-A Modern Introduction}, Cambridge Tracts in Mathematics, \textbf{108}, Cambridge University Press, Cambridge, 2001. 

\bibitem{DAV} E. B. Davies and N. Mandouvalos, ``Heat kernel bounds on hyperbolic space and Kleinian groups", \plms{57}{1988}{102-208}.

\bibitem{HF} E. D'Hoker and D. F. Freedman, ``Gauge boson exchange in $AdS_{d+1}$", \np{544}{1999}{612-632}. 

\bibitem{BW} B. S. DeWitt, \textit{Dynamical Theory of Groups and Fields}, Gordon and Breach, New York, 1965.\\
B. DeWitt, "Quantum field theory in curved spacetime", \pr{19C}{1975}{295-357}.

\bibitem{F} H. Flanders, \textit{Differential Forms with Applications to the Physical Sciences}, Academic Press, 1963.

\bibitem{GA} M. P. Gaffney, ``A special Stoke's theorem for complete Riemannian manifolds", \am{60}{1954}{140-145}. 

\bibitem{GM} S. Giombi, A. Maloney and Xi Yin, ``One-loop Partition Functions of 3D Gravity", \jhp{08}{2008}{007}.

\bibitem{GR} I. S. Gradshteyn and I. M. Ryzhik, \textit{Table of Integrals, Series, and Products}, Academic Press, 5th edition, 1994.

\bibitem{GRI} A. Grigor'yan and M. Noguchi, ``The heat kernel on hyperbolic space", \bul{30} {1998}{643-650}.

\bibitem{GR1} A. Grigor'yan, ``Heat Kernel and Analysis on Manifolds", AMS/IP, Vol. 47, 2009.

\bibitem{HEL} S. Helgason, \textit{Groups and Geometric Analysis}, Academic Press, 1984.

\bibitem{LW} P. Li and J. Wang, ``Weighted Poincar\'{e} Inequality and Rigidity
of Complete Manifolds", \ase{39}{2006}{921-982}

\bibitem{RS} M. Reed and B. Simon, \textit{Functional Analysis}, Academic Press, 1975.

\bibitem{R} S. Rosenberg, ``The Laplacian on a Riemannian manifold", London Mathematical Society Student Texts 31, Cambridge University Press, 1997.

\bibitem{S} R. S. Strichartz, \textit{Analysis of the Laplacian on the complete Riemannian manifold}, \jfa{52}{1983}{48-79}.

\bibitem{JS} J. L. Synge, \textit{Relativity: The general theory}, North-Holland Publishing Company, Amsterdam 1960.

\bibitem{VA} D. V. Vassilevich, ``Heat kernel expansion: user's manual", \pr{388}{2003}{279-360}.

\end{thebibliography}
\end{document}